\documentclass[8.5pt,twoside,twocolumn]{article}

\usepackage[super,sort&compress,comma]{natbib} 
\usepackage{mhchem}
\usepackage{times,mathptm}
\usepackage{sectsty}
\usepackage{balance} 

\usepackage{graphicx} 
\usepackage{lastpage}
\usepackage[format=plain,justification=raggedright,singlelinecheck=false,font=small,labelfont=bf,labelsep=space]{caption} 
\usepackage{fancyhdr}
\usepackage{subfigure}

\newcommand{\bra}{\langle}
\newcommand{\ket}{\rangle}

\begin{document}

\thispagestyle{plain}
\fancypagestyle{plain}{
\fancyhead[L]{\includegraphics[height=8pt]{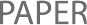}}
\fancyhead[C]{\hspace{-1cm}\includegraphics[height=20pt]{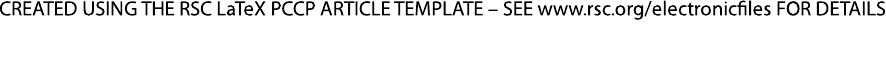}}
\fancyhead[R]{\includegraphics[height=10pt]{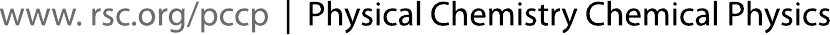}}
\renewcommand{\headrulewidth}{1pt}}
\renewcommand{\thefootnote}{\fnsymbol{footnote}}
\renewcommand\footnoterule{\vspace*{1pt}%
\hrule width 3.4in height 0.4pt \vspace*{5pt}}

\makeatletter 
\renewcommand\@biblabel[1]{#1}            
\renewcommand\@makefntext[1]%
{\noindent\makebox[0pt][r]{\@thefnmark\,}#1}
\makeatother 
\renewcommand{\figurename}{\small{Fig.}~}
\sectionfont{\large}
\subsectionfont{\normalsize} 

\fancyfoot{}
\fancyfoot[LO,RE]{\vspace{-6pt}\includegraphics[height=8.5pt]{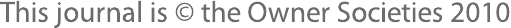}}
\fancyfoot[CO]{\vspace{-6.5pt}\hspace{11.4cm}\includegraphics{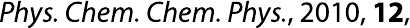}}
\fancyfoot[CE]{\vspace{-6.6pt}\hspace{-12.7cm}\includegraphics{RF.pdf}}
\fancyfoot[RO]{\footnotesize{\sffamily{1--\pageref{LastPage} ~\textbar  \hspace{2pt}\thepage}}}
\fancyfoot[LE]{\footnotesize{\sffamily{\thepage~\textbar\hspace{4.4cm} 1--\pageref{LastPage}}}}
\fancyhead{}
\renewcommand{\headrulewidth}{1pt} 
\renewcommand{\footrulewidth}{1pt}
\setlength{\arrayrulewidth}{1pt}
\setlength{\columnsep}{6.5mm}
\setlength\bibsep{1pt}

\twocolumn[
  \begin{@twocolumnfalse}
\noindent\LARGE{\textbf{Isotopic Effect and Temperature Dependent Intramolecular Excitation Energy Transfer in a Model Donor-Acceptor Dyad}}
\vspace{0.6cm}

\noindent\large{\textbf{Jaykrishna Singh,\textit{$^{a}$} and
Eric R. Bittner$^{\ast}$\textit{$^{a}$}}}\vspace{0.5cm}

\noindent\textit{\small{\textbf{Received 16th February 2010,\newline
Preprint First published on the web 17th February 2010}}}

\noindent \textbf{\small{http://k2.chem.uh.edu/cgi-bin/bib.cgi}}
\vspace{0.6cm}

\noindent \normalsize{We consider here the non-adiabatic energy transfer dynamics for a model bi-chromophore 
system consisting of a perylenediimide unit linked to a ladder-type poly-(para-phenylene) 
oligomer.  Starting from a  semi-empirical parameterization of a model electron/phonon Hamiltonian, 
we compute the golden-rule rate for energy transfer from the LPPP5 donor to the PDI acceptor.  
Our results indicate that the non-adiabatic transfer is promoted by the out-of-plane wagging modes of the 
C-H bonds even though theses modes give little or no contribution to the Franck Condon factors in this system.
We also predict a kinetic isotope effect of $k^{(H)}/k^{(D)} = 1.7 - 2.5$ depending upon the temperature.}
\vspace{0.5cm}
 \end{@twocolumnfalse}
  ]
\section{Introduction}

\footnotetext{\textit{$^{a}$Department of Chemistry, University of Houston, Houston TX 77204-5003}}

Electronic energy
transfer between donor and acceptor units provides the basic energy transport mechanism for 
optical-electronic devices and photosynthetic systems in nature. 
For the case of separated donor/acceptor species, one typically assumes that the
off-diagonal coupling between states, $J$, is independent of the internal vibrational
motions of the two species and that
the internal motions of each species are independent of each other
allowing one to treat them as ``separate baths.''  This {\em ansatz}
is useful since it allows one to compute
transfer rates based upon the spectral overlap between isolated donor
and acceptor states.  
This, along with a other assumptions such as that the coupling can be estimated 
by using the transition dipole moments allows one to write the transfer rate
as 
\begin{eqnarray}
k_{DA} &=& \frac{|J|^{2}}{2 \pi \hbar^2}\int_{0}^{\infty}E_{A}(\omega)
{\cal I}_{D}(\omega )d\omega 
\end{eqnarray}
where $ {\cal I}_{D}(\omega) $ is the fluorescence spectrum of the 
donor  and $E_A(\omega)$ is the (normalized) absorption spectrum of the acceptor.  The coupling
$J$ is the dipole-dipole coupling between transition moments which scales with as $1/R^{6}$ in the separation between donor and 
acceptor species. This is valid only when $R$ is large compared to size of the chromophores themselves.
With in the F\"orster model, energy transfer is
becomes efficient when there is sufficient overlap between the
emission spectrum of the donor and the absorption spectrum of the
acceptor.

\begin{figure}[t]
	         {
	          \centering
	          \includegraphics[width=\columnwidth]{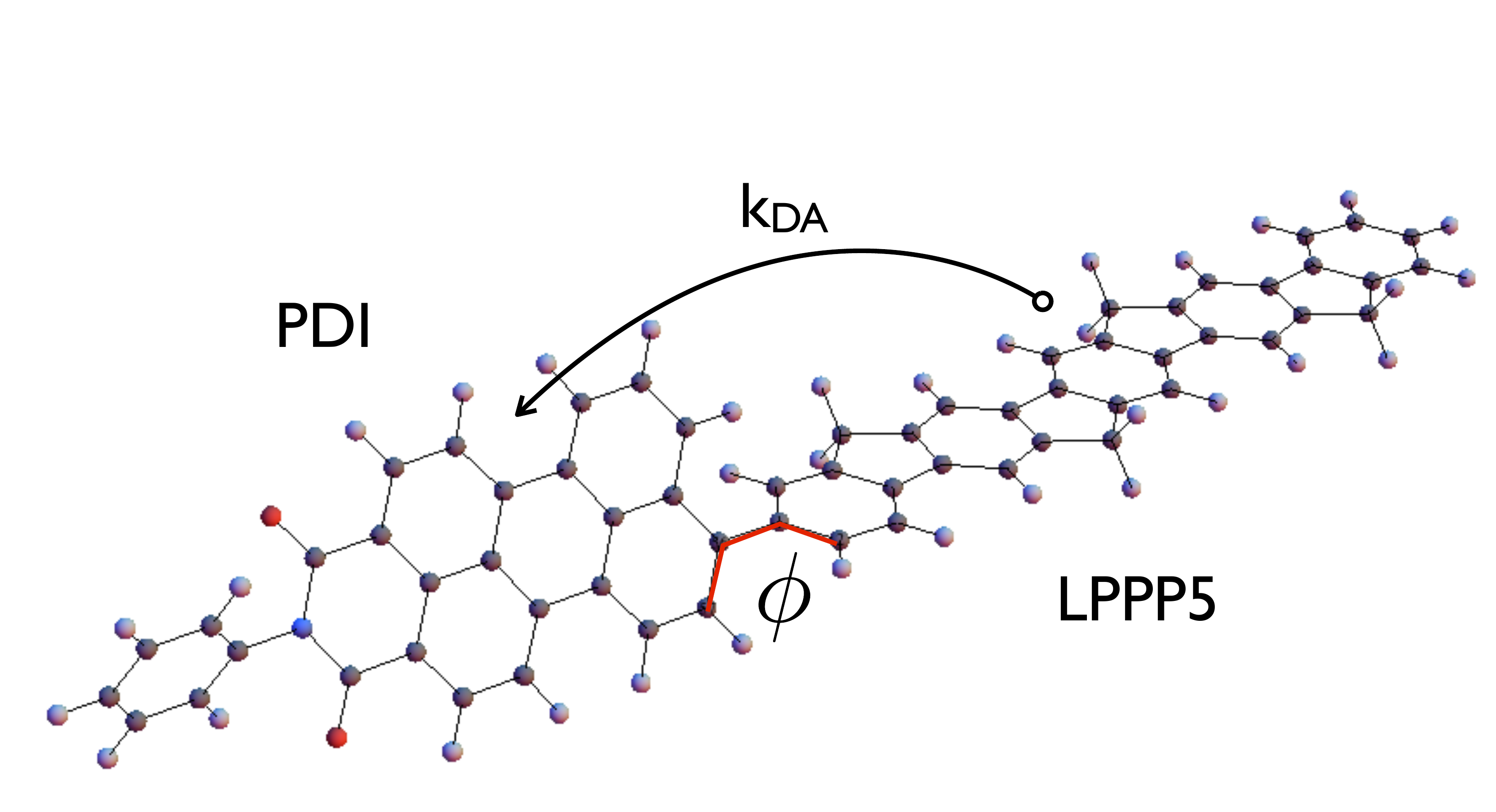}
	       }
               \caption{Chemical structure of donor-acceptor dyad:  LPPP5 (ladder-type poly-({\em p}-phenylene)) PDI (perylenediimide). The highlighted bonds (in red) define the dihedral angle, $\phi$, between the donor and acceptor moieties. For the ground, donor, and acceptor states  $\phi_{gs}= 59.6^\circ$, $\phi_{D}= 44.2^\circ$, and  $\phi_{A}=59.1^\circ$, respectively.
             }
	\label{PDI}
\end{figure}

However, in a number of interesting cases 
energy transfer can be very efficient even in the absence of significant spectral overlap
\cite{hinze:124516,scholes2003,Becker:2005bl,Becker:2006dz}.
This scenario can arise in cases where the surrounding media appears to 
participate in the energy transfer process by retaining memory of the 
donor  state long after the transfer has occurred.   Here, it is 
often useful to invoke within the model a common
``bath'' or ``shared'' sets of modes between the donor and acceptor species
such that the common mode not necessarily appearing in the absorption
or emission spectra participates in the energy transfer event \cite{Beljonne:2009la}.
However, there remains some ambiguity in describing this mode and its
coupling to the electronic transition.  

Our goal in this paper is to devise a systematic approach for determining the types of modes
that participate in phonon-assisted energy transfer process and to
test this approach on a series of donor-acceptor species known to
exhibit efficient energy transfer.  Our starting point will be from a
quantum chemical perspective which will allow us to parameterize our
theoretical description based upon molecular considerations and
systematically improve our description using the appropriate levels of
quantum chemical theory.

\begin{table*}[t]
 \caption{Quantum chemical (AM1/CI(S)) data for LPPP5-PDI dyad.  $E_{n}$ is the 
 vertical excitation energy and $\delta E_{n}$ is the reorganization energy of each state.  $r_{DA}$ is the length 
 of the connecting bond and $\phi$ is the dihedral angle between LPPP5 and PDi moieties for each optimized 
electronic state.}
\begin{center}
 \begin{tabular} {ccllll}
 state  &  designation&  $E_{n}$ (eV)  &   $\delta E_{n}$    (meV)&  $r_{DA}$(\AA)    & $\phi$ \\  
 \hline
$S_{0}$ & DA                 &  0                     &   --                           & 1.38  & 59.6$^\circ$\\
 $S_{1}$ &DA$^{*}$&  2.68 eV         &    82.7  & 1.40       &  44.2$^{\circ}$   \\
$S_{2}$  &D$^{*}$A   &2.98 eV         &    142.6 & 1.38        & 59.1$^\circ$\\
  \end{tabular}
  \hspace{0.1cm}

$S_{2}\to S_{1}$ transition moment (Debye): $\mu_{12}$=0.063 $\hat x$ + 0.078 $\hat y$  +    0.007 $\hat z$ 
 \end{center}
\label{Table1}
 \end{table*}

In this study, we focus on supramolecular donor-acceptor (D-A) unit\cite{b:905753f}, which is made up of a donor molecule ``LPPP5" (Ladder-type poly-(para-phenylene))  and an acceptor molecule ``PDI" (perylenediimide) linked by a single covalent bond. The chemical structure of the D-A unit are shown in Fig.~\ref{PDI}.
Although donor and acceptor are connected by a single covalent bond, quantum chemical 
calculations have shown that  the molecular orientation of LPPPI and PDI are tilted along the single bond approximately
 by $60^\circ$ with respect to each other. 
LPPP and PDI has been extensively studied  and has wide application potential\cite{Wu:2008zr}. LPPP is a p-type electron donating semiconducting polymer which is subject to a particularly low degree of disorder, as the full ladder structure along the backbone blocks chain coiling and bending enforcing a rigid-rod type of geometry\cite{PhysRevLett.91.267403,Wu:2008zr}. 
A rigid coplanar structure of conjugated ladder-type polymers are very suitable for light-emitting applications which enhances the conjugation, carrier mobility, and luminescence intensity\cite{Wu:2008zr}. 
PDI polymers show an n-type, electron conducting behavior and serve as electron-acceptor material\cite{Gunes:2007tg,Singh:2006qv}.
Covalently linked molecular D-A units can be seen as model compounds for D-A blends used in optoelectronic devices. After photoexcitation of D-A unit, excitation energy transfer occurs from an excited donor $(D^*A)$ to an acceptor molecule $(DA^*)$ to a degree determined by dipole-dipole interaction\cite{valeur}.

\section{Theoretical approach}

\subsection{Non-adiabatic Hamiltonian}
For the co-joined species, we can not make a clear separation between donor and acceptor species since there is the possibility of
strong electronic mixing and sharing of vibrational modes.  
Our approach here will be to work within the electronic eigenstate basis and assume that we can treat the electron/phonon coupling as linear in the vibrational coordinates,
so that the entire molecule can be treated within a linearized Born-Oppenheimer approximation. 
We also assume that the normal modes of the ground state are a good description of the modes in the excited states. 
To begin, we shall expand the electronic Hamiltonian about the ground state equilibrium geometry  of the molecule.
\begin{eqnarray}
H(\vec q) = H(0) + \nabla H \cdot \vec q + \frac{1}{2!}\nabla^{2}H q^{2} + \cdots
\end{eqnarray}
Where vector $\vec q$ denotes all the normal mode coordinates of the supramolecular system\cite{b:905753f} 
at the ground state equilibrium geometry $(\vec q =0)$. 
It represents mass-scaled normal coordinate local to the entire donor-acceptor unit with vibrational frequency $\omega$. 
We  compute electronic eigenstates along the $\vec q$ while satisfying 
$$H(\vec q) | \psi_{a} (\vec q) \rangle = E_{a}(\vec q) |\psi_{a}(\vec q)\rangle.$$
Thus, the matrix elements of $H(\vec q)$ can written in a primitive adiabatic basis as
\cite{Longuet-Higgins:1961fk,Longuet-Higgins06241975}
\begin{eqnarray}
H(\vec q) = \left(
\begin{array}{cc}
E_{a}(0) + \vec g_{a}\cdot \vec q   & \vec\nabla_{q} H_{ab}\cdot \vec q \\
\vec\nabla_{q} H_{ba}\cdot \vec q            &  E_{b}(0) + \vec g_{b}\cdot \vec q   
\end{array}
\right)
+ \frac{1}{2}\omega^{2} q^{2}  + \cdots
\end{eqnarray}
Where, $E_{a,b}(0)$ are the vertical energies at the ground state equilibrium geometry $(\vec q =0)$.
$\vec g_{a}$
is the force directed along the vector from the ground state equilibrium geometry  to adiabatic minima 
of excited state $\psi_a(\vec q)$. Similarly, $\vec g_{b}$
is the force directed along the vector from the ground state equilibrium geometry to the adiabatic minima 
of excited state $\psi_b(\vec q)$. We can interpret $\vec g_a$ and $\vec g_b$ as being two non-orthogonal vectors originating at the ground state equilibrium geometry of the supramolecular system and pointing towards the equilibrium geometry of the respective excited states\cite{pereverzev:034104}.

Within a harmonic approximation, we  can determine  the diagonal terms by taking the energy derivative of $E_a$ along a given normal mode direction. 
\begin{eqnarray}
g_{aan} = \frac{d}{dq_n} \langle \psi_a | H | \psi_a \rangle = \left.\frac{d E_a}{dq_n}\right|_{q_n=0} 
\end{eqnarray}
The on-diagonal couplings can be determined numerically from
quantum chemistry as the energy gradient of an excited state energy
taken at the ground-state equilibrium geometry. Thus, $g_{aan}$ are derived from the reorganization energy and shift in the geometry from ground state $(q=0)$ to the excited state ($a$ or $b$).
In the harmonic approximation, the
equilibrium geometry of the molecule in a given electronic excited
state is shifted along each normal mode relative to the ground state
equilibrium geometry by 
$g_{aan}/\omega_n$ with energy
$$\tilde E_a = E_a - \sum_n \frac{g^2_{aan}}{\omega_n^{2}}.$$

The off-diagonal terms, 
$\vec \nabla H_{ab}\cdot \vec q$ are the non-adiabatic couplings between electronic eigenstates $\psi_a(\vec q)$
and $\psi_b(\vec q)$ at the ground state geometry $(\vec q=0)$ belonging to the adiabatic potentials at $E_a(\vec q)$ and $E_b(\vec q)$.
Since $\psi_{a}(\vec{q})$ and $\psi_{b}(\vec{q})$  are electronic eigenstates of $H(\vec q)$ for all $\vec q$,
$\vec\nabla_{q} H_{ab}(q) = 0$.
Consequently, close to the ground-state geometry,  the non-adiabatic coupling  can be rewritten as
\begin{eqnarray}
\vec \nabla_{q} H_{ab} \cdot \vec q &=& \langle \psi_{a} (0)| \vec \nabla _{q}H(0) | \psi_{b}(0) \rangle \cdot \vec q\nonumber  \\
&=& (E_a-E_b)\langle \psi_{a} | \vec \nabla_{q} |\psi_{b} \rangle \cdot \vec q.
\end{eqnarray}
This is often referred to as the ``off-diagonal'' Hellmann-Feynman theorem\cite{habitz:5532}.
We next replace the gradient operator with the phonon momentum operator, $\vec p =- \hbar i \vec \nabla_{q}$.
Then, we recognize that if we write the molecular Hamiltonian in the mass-scaled coordinates as 
\begin{eqnarray}
H_{mol} = \frac{1}{2}\vec p^{2} + H(\vec q)
\end{eqnarray}
then $ [H_{mol}, \vec q] = i\hbar \vec p$. Thus,  $\vec \nabla_{ q } = - [H_{mol},\vec q]/\hbar^{2}$ and we can write
\begin{eqnarray}
\vec \nabla_{q} H_{ab} \cdot \vec q = -\frac{(E_a-E_b)}{\hbar^{2}}\langle \psi_{a} | [H_{mol}, \vec q] |\psi_{b} \rangle \cdot \vec q.
\end{eqnarray}
Next, we assume that $\langle \psi_{a} | p^{2} | \psi_{b} \rangle \approx 0$ 
and can be ignored
so that 
\begin{eqnarray}
\vec \nabla_{q} H_{ab} \cdot \vec q = -\frac{(E_a-E_b)^{2}}{\hbar^{2}e}(\vec \mu_{ab}\cdot \vec q)
\end{eqnarray}
where $\vec \mu_{ab} = e \langle \psi_{a}  | \vec  q | \psi_{b} \rangle$ is the electronic transition dipole 
moment between states $\psi_a$ and $\psi_b$
computed at $\vec q = 0$.  
Here we are using the normal mode coordinates as a general basis for the position operator that can act on 
on the electronic degrees of freedom.   A simple justification for this is that the Hamiltonian $H$ must be in the 
totally symmetric irreducible representation, thus the components of $\vec\nabla H$ must be in the same irreducible representation as $x$, $y$, and $z$ respectively.
Thus, the non-adiabatic coupling can be approximated by taking the projection of the electronic 
coupling between states $\psi_{a}$ and $\psi_{b}$ and projecting this along the displacement vectors for normal mode $\vec q$.
Rewriting $g_{abn}= \vec \nabla H_{ab} \cdot \vec q$ and $\mu_{abn}= \vec \mu_{ab} \cdot \vec q$,
which gives 
\begin{eqnarray}
 g_{abn}  = -\frac{(E_{a}-E_{b})^{2}}{\hbar^{2}e} \mu_{abn}
\end{eqnarray}
as the non-adiabatic (off-diagonal) coupling for $n^{th}$ normal mode.
Hence, the $g_{abn}$ is calculated by projecting the dipole-transition moments $\vec\mu_{ab}$ between excited states
onto the mass weighted normal mode vector $\vec{q}$.  
It is important to recognize that within the molecular/non-adiabatic model, 
nuclear motions that lead to the geometric distortions of the molecule in a given electronic excited state 
{\em may not necessarily} be the same set of modes that couple the two electronic states.  

Let us now write $H$ in terms of phonon operators $q_{n} = \sqrt{\hbar/2\omega_{n}} (a_{n}^{\dagger} + a_{n})$ and 
define $G_{abn} =  \sqrt{\hbar/2\omega_{n}} g_{abn}$.
Here, all the  $G_{abn}$ parameters are in units of energy rather than units of force.
\begin{eqnarray} H &=&\sum_aE_a |a\ket\bra
a|+\sum_{abn} G_{abn}|a\ket\bra b|(a^{\dagger}_n+a_n)
\nonumber \\
&+&\sum_a\hbar\omega_n(a^{\dagger}_na_n+ \frac{1}{2}) \label{Ham} 
\end{eqnarray} 
Here $|a\ket$'s denote
electronic states with vertical energies $\epsilon_a$, $a_n^{\dagger}$
and $a_n$ are the creation and annihilation operators for the normal
mode $n$ with frequency $\omega_n$, and $G_{abn}$ are the coupling
parameters of the electron-phonon interaction which we take to be
linear in the phonon normal mode displacement coordinate. 

We can separate $H$ into a part that is diagonal with respect to the
electronic degrees of freedom,
\begin{eqnarray} H_0&=&\sum_aE_{a} |a\ket\bra a| +\sum_{an}G_{aan}|a\ket\bra
a|(a^{\dagger}_n+a_n) \nonumber \\
&+&\sum_n\hbar\omega_n(a^{\dagger}_na_n + \frac{1}{2}) \label{H0}
\end{eqnarray} 
and an off-diagonal part $V$ 
\begin{eqnarray} V={\sum_{abn}}'G_{abn}|a\ket\bra
b|(a^{\dagger}_n+a_n), \label{V} 
\end{eqnarray}
 where the prime at the summation sign indicates that the terms with
$a=b$ are excluded.  This separation is useful for the following two
reasons. First, in many systems only off-diagonal coefficients
$G_{abi}$ are small compared to $G_{aan}$.
Hence, $V$ can be treated as a perturbation. Second, for many cases of
interest, the initial density matrix commutes with $H_0$. In this
case, the separation gives simpler forms of the master equations.

\subsection{Obtaining model parameters from quantum chemisty}

One of the advantages of our approach is that one can in principle arrive at a complete parameterization 
of our model Hamiltonian from quantum chemical considerations.  
For the LPPP5-PDI di-chromophore unit shown in Fig.~\ref{PDI} we used
the semi-empirical AM1/INDO model as implemented in the AMPAC package\cite{Dewar:2002jt, ampac}. 
This approach is robust for system such as this and gives reliable energetics and geometry with a reasonable 
amount of computational overhead.   
We first performed ground-state optimization and normal mode analysis
and then performed  configuration interaction (CI) calculations to obtain the optimized geometries 
of the two lowest singlet electronic excited states. 
We also obtained the transition dipole moment between excited states using this procedure.
Relevant data from these calculations are presented in Table~\ref{Table1}.

\begin{figure}[t]
\includegraphics[width=\columnwidth]{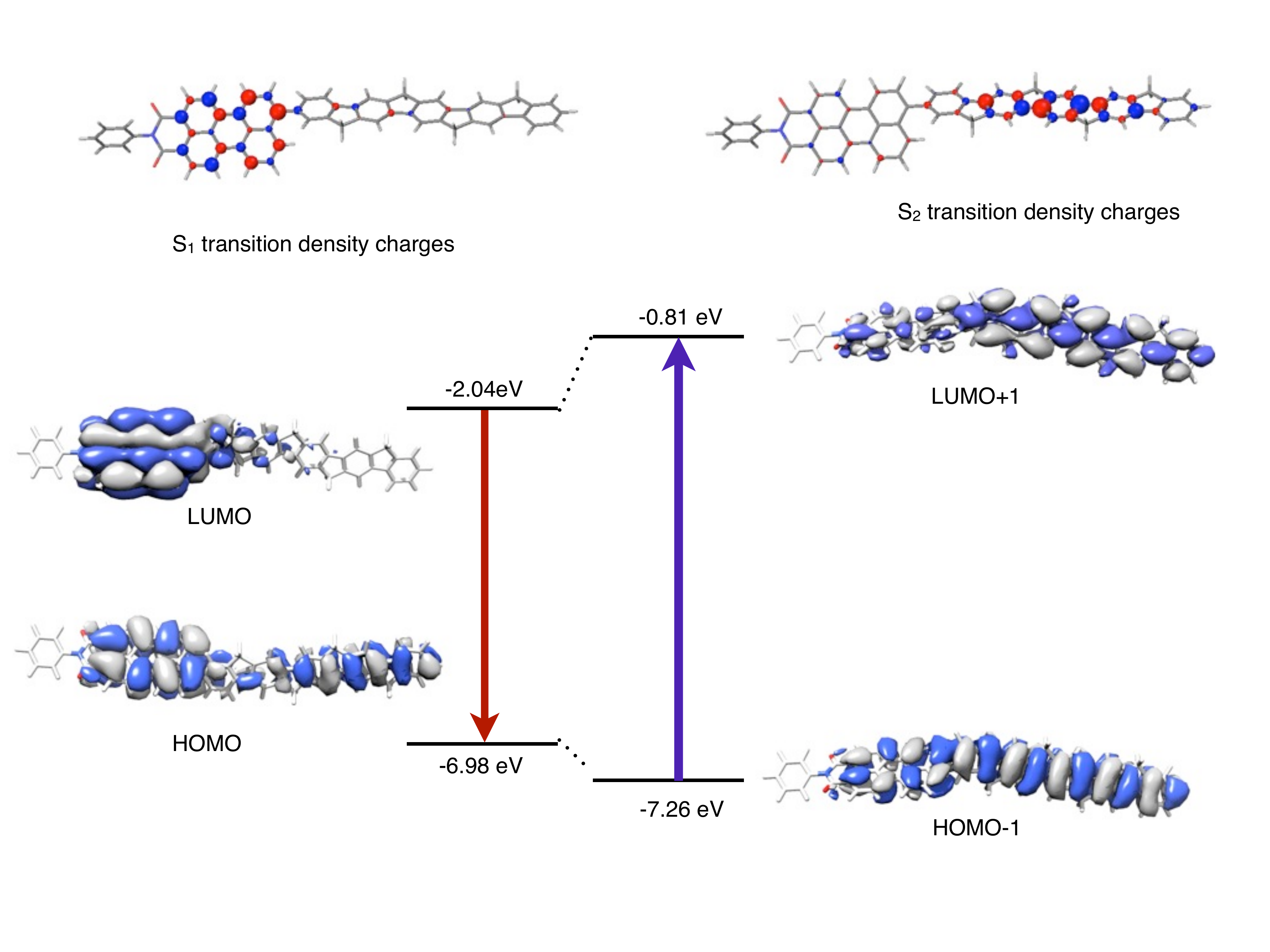}
\caption{Molecular orbitals and energies of the co-joined LPPP5-PDI  system along with the transition charges
for the $S_{o}\to S_{1} $ and $S_{o}\to S_{2}$ 
transitions.}
\label{homo-lumo}
\end{figure}

\begin{figure}[h]
	\subfigure[]{\includegraphics[width=0.49\columnwidth]{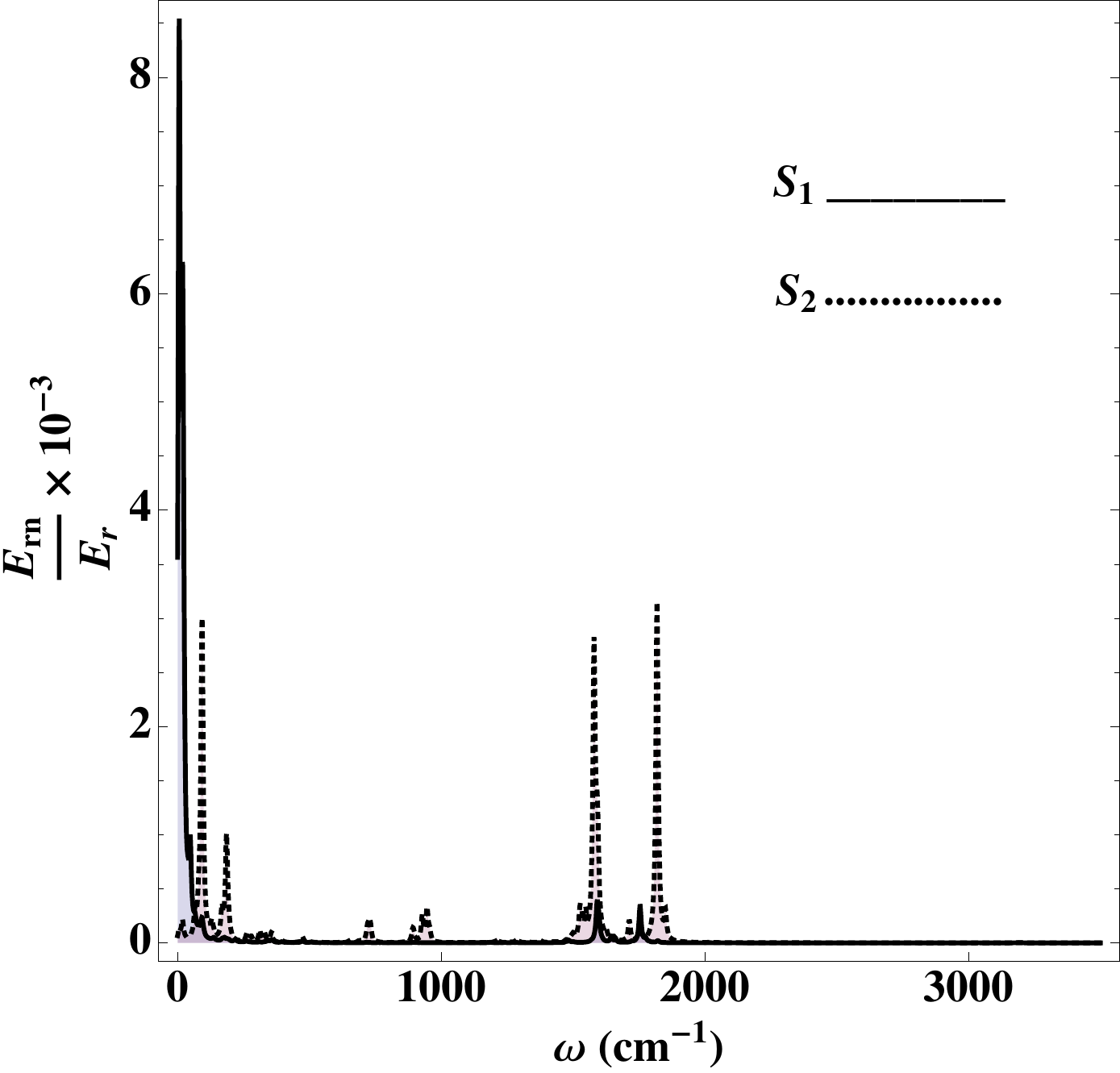}}
	\subfigure[]{\includegraphics[width=0.49\columnwidth]{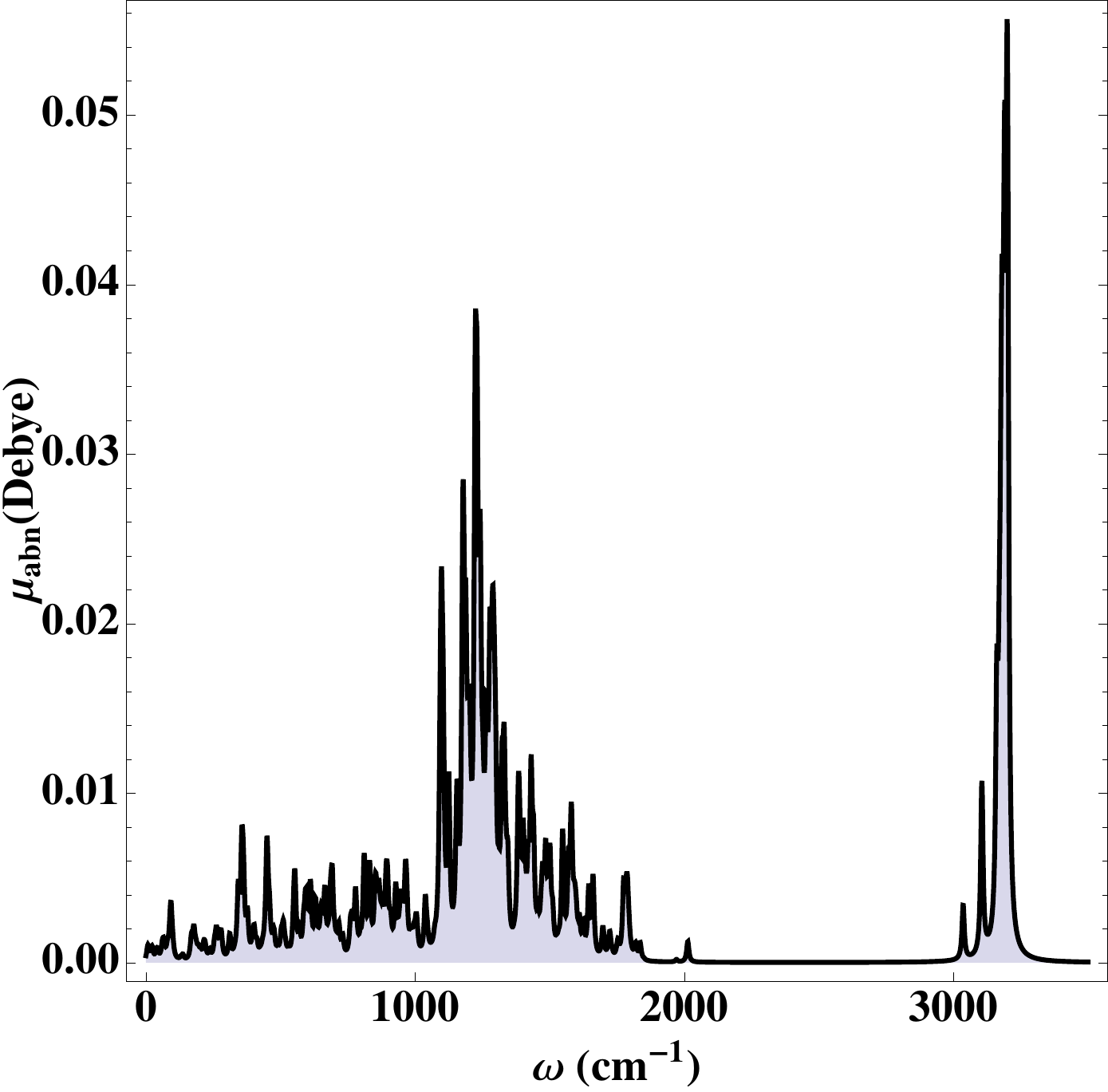}}
		\caption{
		(a)
		Contribution from each normal mode to the reorganization energy of the donor (black: solid line)
		 and acceptor (dotted line) states for the LPPP5-PDI di-chromophore system.
		(b) Magnitude of transition dipole moment vector $\vec \mu_{ab}$ as projected on to each normal mode 
		 between states $S_{1}$ and $S_{2}$ }
	\label{lpp5data}
\end{figure}

For LPPP5-PDI dyad, the $S_{2}$ donor state lies 2.98 eV above the ground state and is largely localized on the LPPP5 end of the molecule.  The $S_{1}$ (acceptor) state is lower at 2.68 eV.  Both are optically coupled to the ground states. 
If the system were forced to be 
planar, the $\pi$-conjugation would extend across the entire molecule.  
The two components of this molecule
 are linked by a bond connecting C atoms participating in the $\pi$ conjugation of both moieties. 
 In the ground and relaxed $S_{2}$ states, the dihedral angle between conjugated domains is 
nearly 60$^\circ$.    In the $S_{1}$ (acceptor) state the C-C bond connecting the two moieties increases slightly 
to 1.40 \AA\ and the system becomes slightly more planar ($\phi  = 44.2^{\circ}$).  
In the electronic ground state, the dihedral angle formed by the respective molecular planes is close to 60$^{\circ}$ suggesting that
the $\pi$ system should be localized on the donor and acceptor moieties. 
 In Fig.~\ref{homo-lumo} we show the HOMO-1 through LUMO+1 molecular
orbitals for the dyad along with the associated transition densities for the $S_{o}\to S_{n}$ transitions.  By and large, the HOMO and LUMO
orbitals are localized on the PDI side of the dyad, although there is a significant amount of $\pi$ amplitude leaking though the linkage over to the LPPP5 end.   
Likewise, the LUMO-1 and HOMO+1 are mostly localized on the LPPP5 end, but one can clearly see significant leakage over to the PDI end.  

At the top of Fig.~\ref{homo-lumo} we show the  corresponding transition densities for the $S_o \to S_{1}$ (2.68 eV) and  $S_o \to S_{2}$ (2.98 eV) transitions.   Since the AM1 model is based upon the zero-differential overlap approximation, the transition densities are represented as 
atom centered charges rather than spatial densities.  The transition moment is obtained by multiplying the local charge by the atomic coordinate vector
and summing over all atoms.   By and large, the vertical transitions are localized to the respective donor and acceptor sides of the molecule with some
``leakage'' across the covalent linking bond to the other side of the molecule.  This weak $\pi$-communication could in part account for the efficient 
non-adiabatic relaxation in this system.

\begin{figure}[t]
\includegraphics[width=\columnwidth]{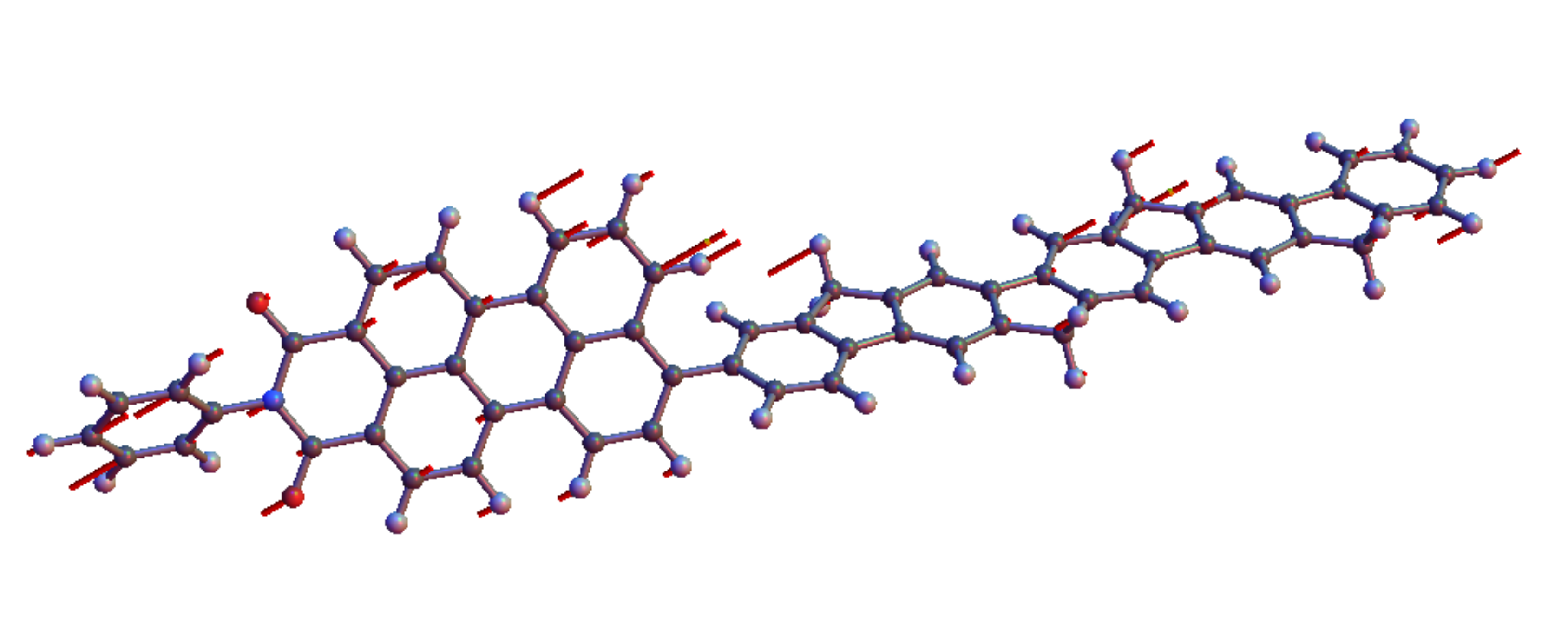}
\caption{Projection of DA transition moment,  $\vec\mu_{DA}$, onto atomic displacements for PDI-LPPP5 dyad.}\label{project}
\end{figure}

In Fig.~\ref{lpp5data}a we show the contribution to the total adiabatic 
reorganization energy from each normal mode for both the donor and acceptor
states.   As discussed above, these are proportional to the magnitude of the diagonal contribution to the electron/phonon  coupling.  
 For the $S_{1}$ state, the largest contribution to the reorganization energy comes from a set of low frequency modes (c.a. 30 cm$^{-1}$). 
 These correspond to the frustrated dihedral torsional  motion at the linking point between the two parts of the molecule.  
This feature is not present in the coupling spectrum for the other state.  Common to both, however, are 
contributions from modes around 1600 cm$^{-1}$ and 1800 cm$^{-1}$.  These correspond to the C=C bond stretching modes and are consistent 
with the the vibronic fine-structure peaks observed in most conjugated polymer systems\cite{karabunarliev:11372,karabunarliev:5863}.
 
Fig.~\ref{lpp5data}b shows the projection of the transition moment between the $S_{1}$ and $S_{2}$ states onto 
the normal  modes of the system.  As noted above, these are related contributions 
to the non-adiabatic couplings between the two states. 
Here we note two distinct contributions to the non-adiabatic coupling.   These correspond to the wagging (at 1200 cm$^{-1}$ ) 
and stretching (at 3100 cm$^{-1}$) of C-H bonds attached to the conjugated rings of the system.  In Fig.~\ref{project}
we illustrate this graphically by drawing the projection of the transition moment on to the individual atomic displacements of the system. The relative length of each vector indicates the relative component of particular atomic displacement along the total transition moment when summed over normal modes.   Surprisingly, the modes that contribute strongly to the 
adiabatic reorganization of each state give little contribution to the non-adiabatic coupling between the two states. 
This is surprising since one expects  that the states with the largest electron-phonon coupling  would give both the 
largest contribution to both the reorganization and the state-to-state transitions.

\begin{figure}[h]
	         {
	          \centering
	          \includegraphics[width=\columnwidth]{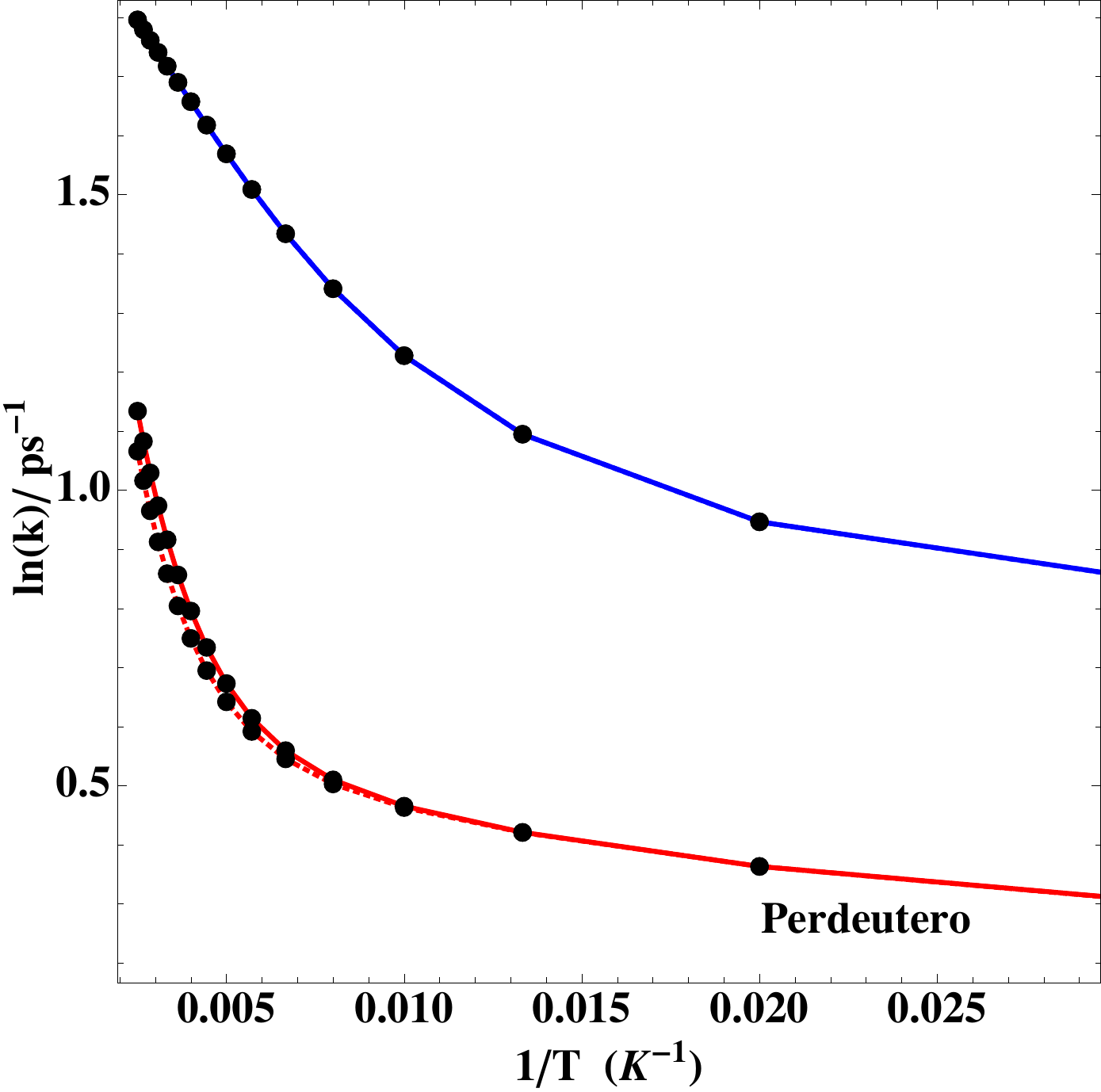}
	       }
               \caption{
               Arrhenius plot of $\ln k (T)$ versus ${1}/{T}$ comparing hydrogen and deuterium substituted donor-acceptor system. Solid line represents the contribution from off-diagonal electron-phonon couplings $\{g_{abn}\}$ of all the vibrational mode between eigenstate $a$ and $b$ which have frequencies in the range of $5-3200 cm^{-1}$, 
and dotted line represents the contribution from off-diagonal electron-phonon couplings of those vibrational modes which have the frequencies  $\leq 2000$ $cm^{-1}$ i.e. dotted line: $g_{abn} \leq 2000$ $cm^{-1}$. }
	\label{lnk}
\end{figure}

\begin{figure} [t]
\subfigure[ ]{\includegraphics[width=0.48\columnwidth]{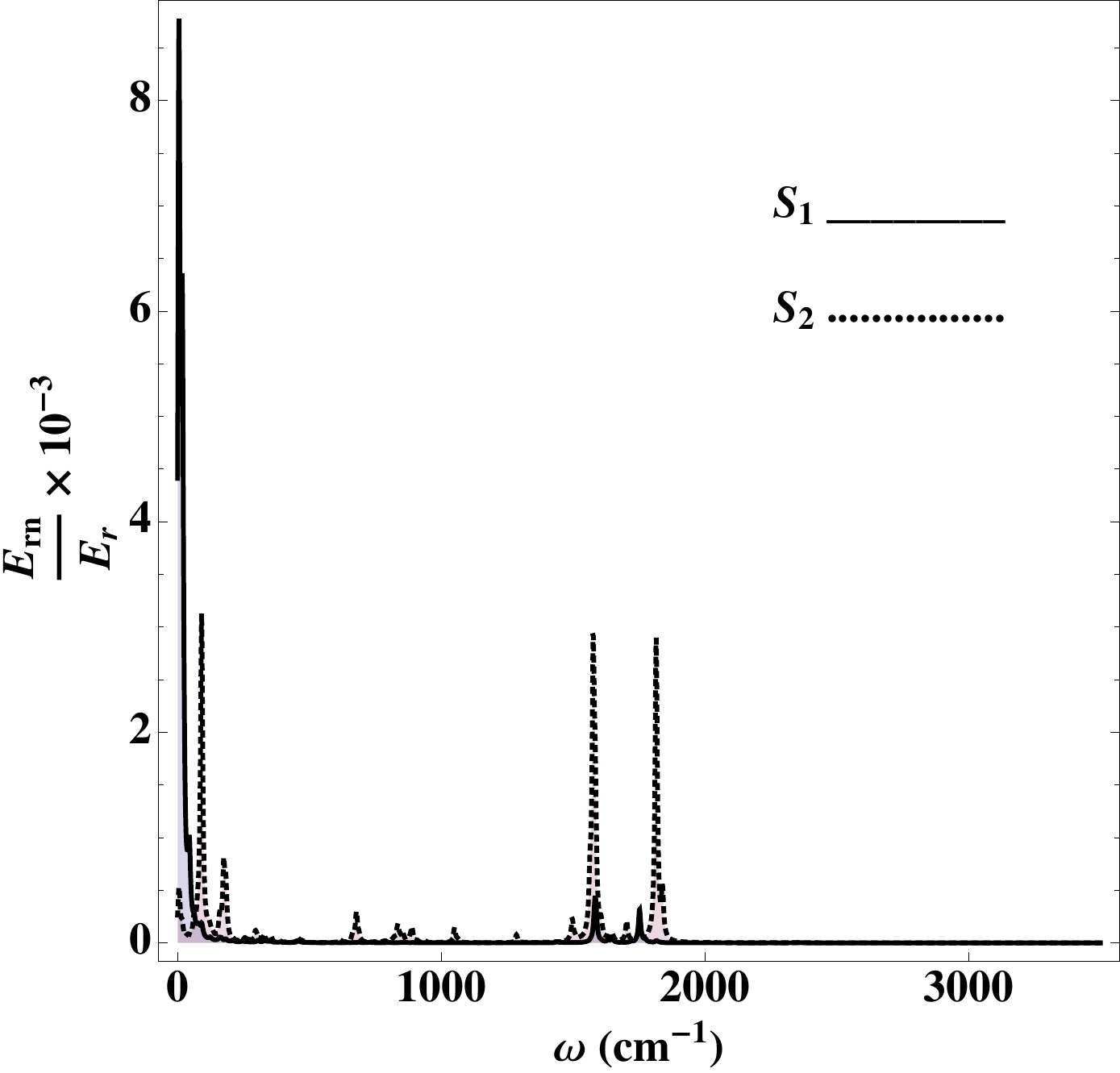}	}
\subfigure[ ]{\includegraphics[width=0.48\columnwidth]{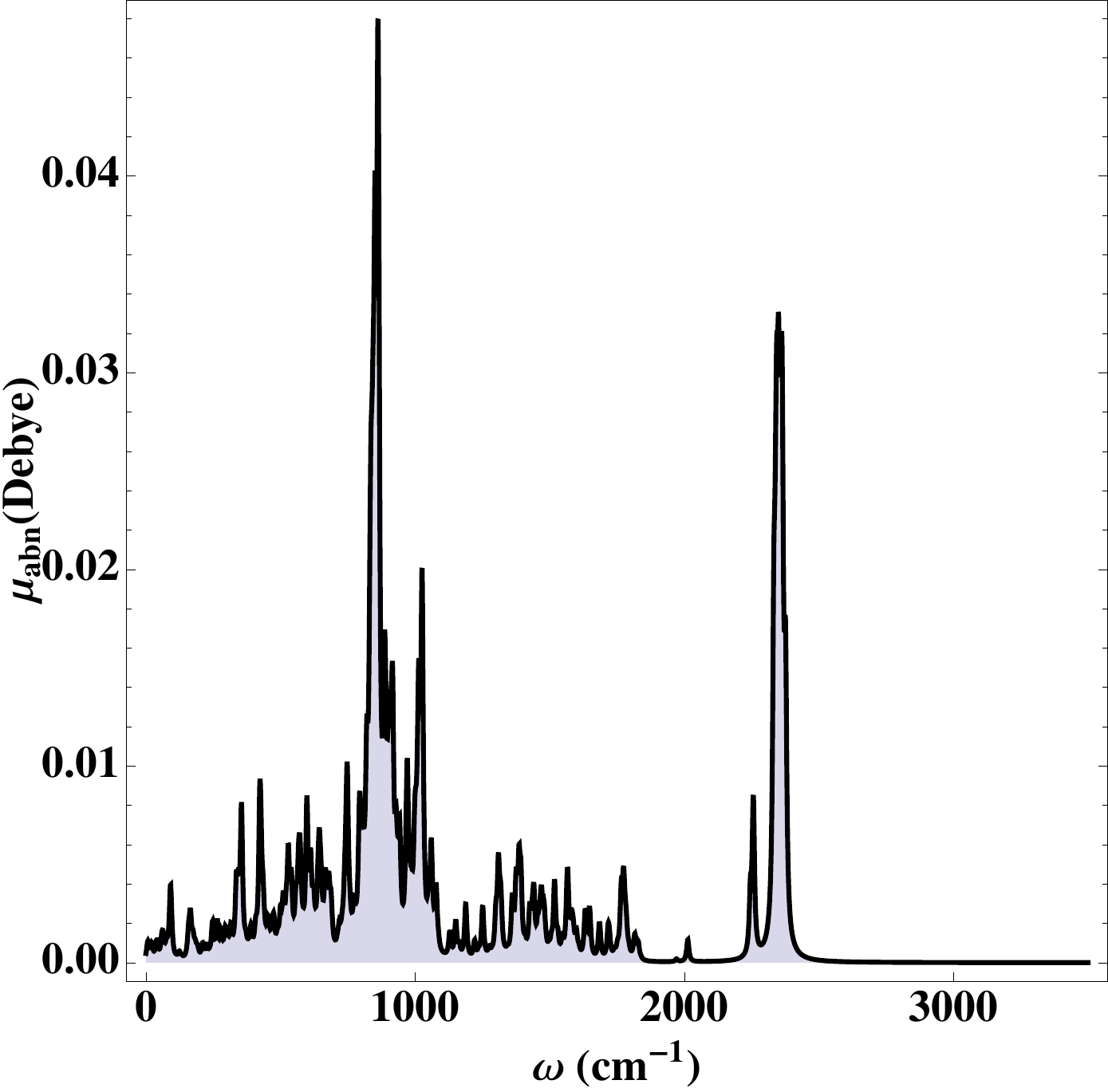}	}
\caption{ Diagonal (a) and off-diagonal couplings (b) as in  Fig~\ref{lpp5data} except for fully deuterated species}
\label{perdeutero}
\end{figure}

\section{Non-adiabatic relaxation rates}

Having parameterized our model,  we can move on to compute the electronic energy transfer rate between the LPPP5 and PDI  parts of the molecule.  
We now take the off-diagonal (non-adiabatic) electron-phonon coupling terms as the weak perturbation
in order to describe transitions between the electronic eigenstates. 
In Ref.\cite{pereverzev:104906} Pereverzev and Bittner developed a time-convolutionless 
approach for treating non-adiabatic transition for systems described by the Hamiltonian given above. 
We begin by performing the polaron transform using
\begin{eqnarray}
U&=&e^{-\sum_{an}\!\!\frac{G_{aan}}{\hbar\omega_n}|a\ket\bra a|
(a^{\dagger}_n-a_n)}\nonumber \\
&=& \sum_{a}|a\ket\bra a|e^{-\sum_{n}\!\!\frac{G_{aan}}{\hbar\omega_n}(a^{\dagger}_n-a_n)}
\label{unitary} 
\end{eqnarray} 
in which our transformed Hamiltonian becomes
\begin{eqnarray} 
\tilde H_0&=&U^{-1}H_0U \\
&=&\sum_a\tilde E_{a} |a\ket\bra a|  \nonumber \\ 
&+&\sum_n\hbar\omega_n(a^{\dagger}_n+\frac{G_{aan}}{\hbar\omega_{n}})(a_n 
+ \frac{G_{aan}}{\hbar\omega_{n}}).
 \end{eqnarray}
where the renormalized electronic energies are 
\begin{eqnarray}
\tilde E_{a}=E_a-\sum_{n}\frac{G_{aan}^2}{\hbar\omega_n}.  
\end{eqnarray}
Applying the same unitary transformation to $V$ gives 
\begin{eqnarray}
\tilde{V}=\sum_{abn} |a\ket\bra b|\hat M_{abn}
 \label{opm}, 
\end{eqnarray}
where the system-bath operators are
\begin{eqnarray} \hat M_{abn}=G_{abn}\left(a^{\dagger}_n+
a_n-\frac{2G_{aan}}{\hbar\omega_n}\right)e^{\sum_{m}\frac{(G_{aam}-G_{bbm})}{\hbar\omega_m}(a^{\dagger}_m-a_m)}
\label{opm}.
\end{eqnarray}
At this point it is useful to connect the various terms in our Hamiltonian 
with specific physical parameters. 
The terms involving $(G_{aan}-G_{bbn})/\hbar\omega_n$ 
can be related to the reorganization energy
$$
E_{r} =\sum_n \frac{(G_{aan}-G_{bbn})^2}{\hbar\omega_n} = \sum_n \hbar\omega_n S_n
$$
where $S_n$ is the Huang-Rhys factor for mode $n$ which is related to the Franck-Condon factor
describing the overlap between the $v_n=1$ vibronic state in one electronic state
with the $v_n=0$ vibronic state in the other. 
Likewise, the energy difference between the renormalized energies is related to the 
driving force of the transition,
$$
\Delta E_{ab} = \tilde E_{a}-\tilde E_{b}.
$$
In the transformed picture the
electronic transitions from state $|a\ket \to |b\ket$ are accompanied
not only by the creation or annihilation of a single phonon of mode
$n$ but also by the displacements of all the normal modes.  
Once we have transformed to the shifted (polaron) representation, we can construct the 
golden-rule rate in terms of the autocorrelation function
\begin{eqnarray}
k_{ab} &=& 2 {\rm Re}\int_0^\infty d\tau 
C_{ab}(\tau)  e^{-i(\tilde\epsilon_b-\tilde\epsilon_a)t}   ,
\label{clgr}
\end{eqnarray}
where $k_{ab}$ is the non-adiabatic rate ($k_{NA}$), and $C_{ab}(\tau) = \langle\tilde M_{a b}(0) \tilde M_{ b a}(\tau)\rangle_{th}$ is the autocorrelation of the  
polaron transformed electron-phonon operators
\begin{eqnarray}
C_{ab}(\tau) &=&
\left\{
\left[
\sum_{n} G_{a b n}
\left( \Delta_{abn}(\bar{n}_n+1)e^{i \omega_n \tau}  - \bar{n}_n e^{-i \omega_n \tau})+ \Omega_{abn}\right)     
\right]^{2}\right.
\nonumber  \\
&+& \left.\sum_{n} G_{abn}^{2}\left[(\bar{n}_m+1)e^{i \omega_n \tau} +\bar{n}_ne^{-i \omega_n \tau}) \right] \right\} 
\nonumber \\
&\times& q_{a b}(\tau) f_{a b}(\tau),
 \label{correl}
\end{eqnarray}
where
\begin{eqnarray}
\Delta_{abn} = \frac{G_{aan}-G_{bbn}}{\hbar\omega_{n}},\\
\Omega_{abn}= \frac{ (G_{a a n}+ G_{b b n})} {\hbar\omega_n},
\end{eqnarray}
and
\begin{eqnarray}
\bar{n}_n = \frac{ 1}{ \exp (\hbar \omega_n / k_B T) -1},      
\end{eqnarray}
is the Bose-Einstein occupation of the $n$th phonon mode.

Note that in the first term in Eq. ~\ref{correl} gives the non-Condon contributions to the correlation function.
The two time dependent quantities are given by  
\begin{eqnarray}
q_{a b}(\tau)= e^{i \sum_n \left(\frac{G_{a a n}-G_{b b n}}{\hbar\omega_n}\right)^{2} \sin(\omega_n \tau)}  ,
\end{eqnarray}
and
\begin{eqnarray}
f_{a b}(\tau) = e^{-2 \sum_n(\bar{N}_n+\frac{1}{2}) \left(\frac{G_{a a n}-G_{b b n}}{\hbar\omega_n}\right)^{2}(1-\cos(\omega_n \tau))}.
\end{eqnarray}
These are related to the spectral density of the diagonal terms and as such are related to the 
Franck-Condon overlap integrals between the phonon modes in the $a$ and $b$ states. 
The decay of $ \langle\tilde M_{a b}(0) \tilde M_{ b a}(\tau)\rangle_{th} $ is related to the decoherence time 
between the two electronic states and we have found that for the system at hand, the golden rule limit is clearly reached in 30 fs for the lowest temperatures  considered in this study. 
Because our formalism does not include dissipative terms, there is a finite recursion time for the correlation function. 
However, for larger systems this recursion time is very long and we assume that all correlation functions vanish for all time 
once the initial decay has occurred.  

In Fig. ~\ref{lnk} we show an Arrhenius plot of the non-adiabatic relaxation rates ($k_{NA}$)
between the LPPP5 and PDI.  
At T =150K, $k_{NA} = 4.4 {\rm ps}^{-1}$ which is consistent with energy transfer rates observed in related systems.
As $T\to 0K$, $k_{NA} \to 1.6 {\rm ps}^{-1}$ reflecting the contribution of zero-point  vibrational motion. 
At higher temperature, we see $\log k_{NA} \propto 1/T$ 
and conclude that the  energy transfer is  activated process with $E_{A} =7.56 {\rm meV}$.  
In order to assess the relative contribution to the non-adiabatic rate from the two types of coupling modes identified in 
Fig.~\ref{lpp5data}, we performed a series of rate constant calculations in which we limited the non-adiabatic couplings to include only modes above or below a 2000 cm$^{-1}$ cut-off.  This cut-off lies clearly between the two peaks seen in the coupling spectra.  Over the entire thermal range, there is no quantitive difference between the fully coupled model and a model including only couplings from the C-H wagging modes.

 \begin{table*}[t]
 \caption{Kinetic data for energy transfer in the high-temperature limit $(E_a)$.  
 }
 \begin{center}
 \begin{tabular} {lcccc }  
             &
               \multicolumn{2}{c}{LPPP5-PDI-$h$}&
               \multicolumn{2}{c}{LPPP5-PDI-$d$}   \\  \hline 	
               \hline
    coupling           &  $E_{A}$ (meV)  & $k(150K)$ ps$^{-1}$  &     $E_{A}$ (meV)  &  $k(150K)$ ps$^{-1}$   \\ \hline
    $g_{abn}$      &    7.69      &4.19   &    15.92  & 1.74   \\  \hline
 $2\times g_{abi}$   &    7.69   & 16.77     &    15.92  & 6.99       \\  \hline
$0.5 \times g_{abi} $        &    7.69     & 1.04    &  15.92  & 0.43          \\  \hline
$ g_{abi} $ cut-off below $2000$ $cm^{-1}$  &   7.68 & 4.19   &  14.61 & 1.72
\end{tabular}
\end{center}
\label{Table2}
 \end{table*}

\subsection{Kinetic isotope effect}

Having identified that C-H wagging modes drive the non-adiabatic energy transfer from the LPPP5 to PDI parts of the molecule, $k_{NA}$ should be sensitive to isotopic substitution of H for D. 
There are two factors contributing to an isotope effect. First,  the coupling strengths themselves depend upon the 
phonon frequency through $G_{abn} = g_{abn}\sqrt{\hbar/2\omega_{n}}$.   Since the $g_{abn}$ do not depend upon the phonon frequency,  we can roughly estimate that $G_{abn}^{(H)} \approx 2^{-1/4}G_{abn}^{(D)}$ for modes involving 
proton motions.  Since the correlation function and hence the golden-rule rate goes as $G_{abn}^{2}$,
 would give a factor of $\sim\sqrt{2}$ to the overall rate.  
 The other contributions to the rate stem from the Franck-Condon terms arising from the 
 displacement between the  two adiabatic potentials.   Upon isotopic substitution, there will be shift in the 
 zero-point energies of each state and generally, the nuclear wave function will be more localized.
  For the case at hand, the normal modes involving the proton motions give nearly all the contributions to the off-diagonal 
 coupling ($G_{abn})$ but contribute little to the Franck-Condon factors.
 Consequently, we can make a ``back of the envelope'' estimate that  $k^{(H)}/k^{(D)}$ is approximately
 $\sqrt{2}$ for systems where non-adiabatic transitions are mediated by proton motions.

To test this, we determined the normal modes for perdeutero-LPPP5-PDI  system in which all 35 protons were replaced by deuterons.  
Since the electronic transition moment and nuclear geometries for the ground and excited states
do not depend upon the nuclear mass, we can use this data to construct a non-adiabatic model for energy transfer for the 
deuterated system. 
The coupling spectra for the perdeuterated system are shown in Fig.\ref{perdeutero}.  
Comparing to the perprotonated system, 
the $g_{aan}$ diagonal coupling terms are virtually identical.  This is not surprising since the $g_{aan}$ and $g_{bbn}$ 
terms in this system reflect C=C bond distortions and do not involve the C-H modes.  
This suggests that the Franck-Condon factors between the donor and acceptor states are more or less identical 
for the perproto- and perdeutero- molecules.
On the other hand for the off-diagonal non-adiabatic terms, $g_{abn}$, 
there is a systematic shift towards lower frequencies of all
the major peaks in  Fig.~\ref{perdeutero}a relative to those in Fig. \ref{lpp5data} 
reflecting the effect of isotopic mass on the C-H frequencies.($\omega_{H} \approx \sqrt{2}\omega_{D}$).

\begin{figure}[t]
\includegraphics[width=\columnwidth]{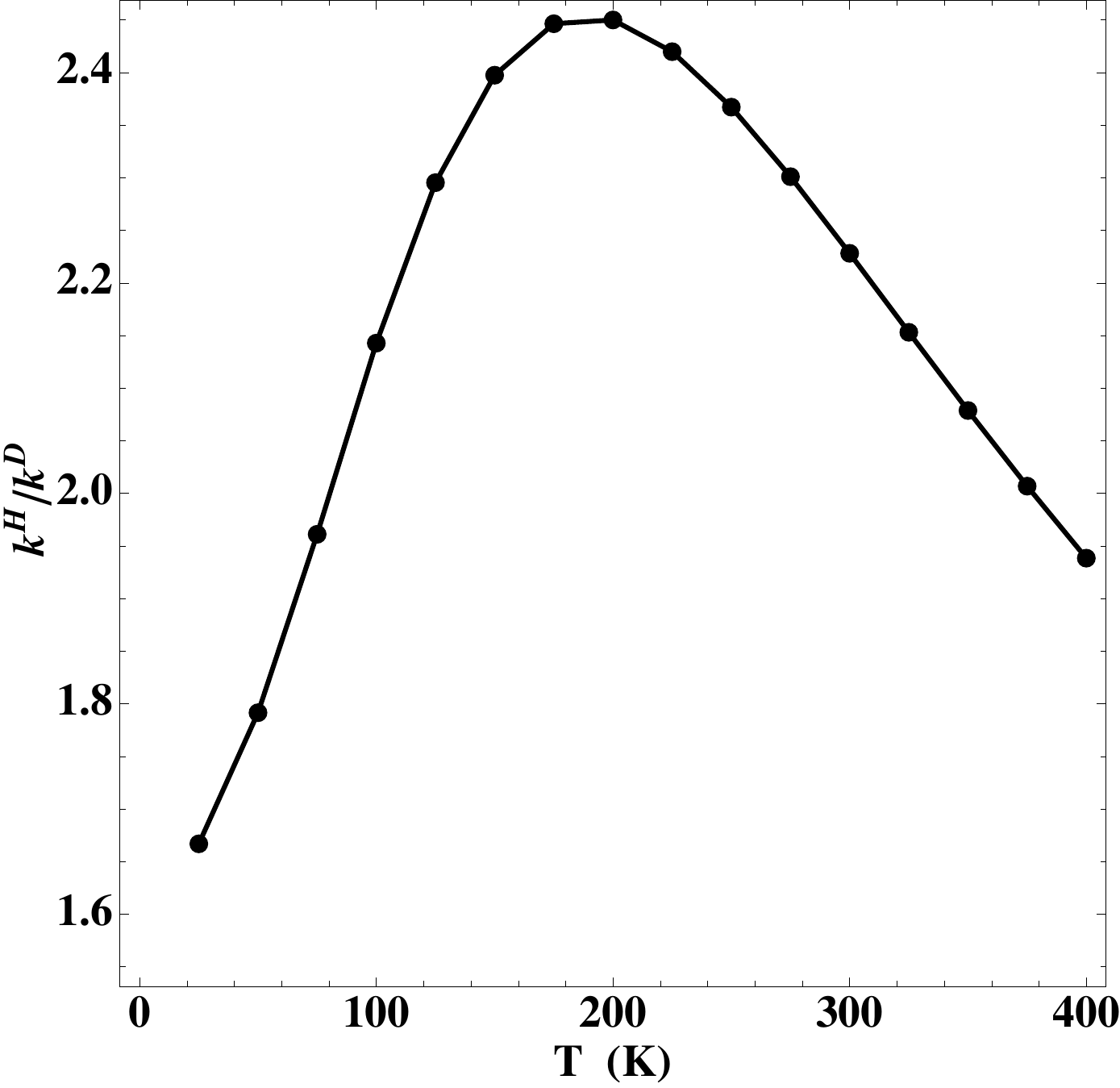}
\caption{Ratio of non-adiabatic transition rates $(k^{(H)}/k^{(D)})$ over the entire temperature range considered herein. }\label{compare}
\end{figure}

In Fig.~\ref{lnk} we show the kinetic data for the fully deuterated LPPP5-PDI-$d_{35}$  and in 
Fig.~\ref{compare} we plot $k^{(H)}/k^{(D)}$ over the 25K to 400K temperature range.  
Over the entire temperature range considered here, $k^{(H)}/k^{(D)}> \sqrt{2}$ which is consistent 
with our estimate above. 
 In the high-temperature limit $(T>250K)$,  we can estimate the Arrhenius activation energy
as in the LPPP5-PDI-$h_{35}$ case.  
Comparing the $E_{A}$ for the perproto- and perdeutero-cases, $E_{A}^{(D)} \approx 2 E_{A}^{(H)}$.
In Table ~\ref{Table2} we given a summary of the activation energies for 
various cases.   Doubling or reducing the off-diagonal couplings by a factor of 2 has little effect on the activation energy
indicating that that even if there were inaccuracies in the couplings themselves, the overall energy transfer kinetics is
determined by the proton vibrational frequencies.  Finally, we considered the effect of truncating the 
coupling spectrum to include {\em only} the C-D wagging modes.  As shown in Fig.~\ref{lnk}, there is some
sensitivity to the overall rate at higher temperature, but overall the kinetics are insensitive to this cut-off indicating that
the C-D modes are the coupling modes in this case.

\section{Discussion}

In this paper we have presented a methodology for determining non-adiabatic electronic transitions in large multi-chromophore systems starting from a common quantum chemical description.   We believe this
 approach will be particularly useful for studying systems in which one can not make a clear separation between the donor and acceptor species and in cases where the are common sets of vibrational motions.   One crucial theoretical observation that comes out of this study is that in systems such as this, nuclear motions that 
 are involved in the geometric reorganization of the molecule in a given excited state may not necessarily be the same set of motions that are involved in the coupling between excited states.  In fact for the system at hand,  we identified that the 
 C-H wagging motions couple the transition between the D and A state and not the C=C modes or ring torsional  modes, 
 even though these modes are involved in the geometric relaxation of the two states.  
We can understand this coupling in the following way:   In the absence of phonons, the two states are coupled via a transition  moment, $\mu_{ab}$. Consequently, a photon with  polarization along $\mu_{ab}$ induces an oscillation of the electronic state (of which the transition density would provide a snap-shot).  In the same sense, in an IR transition the transition intensity is determined by the extent to which the permanent dipole moment of the molecule is modulated by a given normal mode.  Consequently, those modes  with the largest projection along the electronic transition moment will provide an oscillating electric field polarized in the right way to promote the transition.  This is analogous to the situation in resonance Raman spectroscopy where  vibrational modes along the transition dipole are also coupled to the photon field and can  participate in the electronic transition.  Our identification of the nonadiabatic coupling vector with the 
state to state transition dipole moment is best considered a ``propensity'' rather than a selection rule.  None the less it suggests that modes that are in the same irreducible representation as the dipole operator will give the strongest contribution to the non-adiabatic coupling. 

We also predict that in system such as this that the should be a clear kinetic isotope effect over the entire temperature range. 
While kinetic isotope effects are well known to occur in H transfer reactions\cite{Anslyn:2006hp}.  In this case, changing the mass of the reacting atom modifies the zero-point energy and hence increases the activation energy required for bond 
cleavage.  Here, we propose that the non-adiabatic  energy transfer rates will be sensitive to nuclear mass.  
This effect is most analogous to non-adiabatic relaxation dynamics of the solvated electron.  
In this case, ultrafast pump-probe experiments by Barbara's group indicate a $k^{(H)}/k^{(D)} = 1.4$ in comparing the 
ground-state recovery of an excess electron in water versus heavy water\cite{Yokoyama:1998cy}.   The effect was largely 
attributed to the quantum mechanical librational modes of the surrounding solvent cavity\cite{PhysRevLett.66.3172,schwartz:6997,bittner:8611,prezhdo:5863,borgis:064501}.   

The close analogy between non-radiative transition and energy transfer events came be traced to early works by Robinson and co-workers \cite{robinson:1962,robinson:1187,breuer:3615,burland:4548,Lin:1971fk} and kinetic isotope effects in non-radiative relaxation rates for various deuterated aromatic systems has been known since the 60's\cite{breuer:3615,Lin:1971fk}.  For example, deuteration dramatically increases the triplet lifetime of various deuteronapthalene systems 
as observed in ESR experiments by Hutchenson\cite{hirota:1561}.  Similar effects have been reported for the 
$^{3}B_{1u}\to\ {^{1}A}_{1g}$ triplet to singlet conversion in benzene and benzene-$d_{6}$ with $k^{(H)}/k^{(D)}$
ranging between 2 and 4 \cite{li:605}.  For example, in Ar matrix at 4K, 
$k^{(H)}/k^{(D)} = 5.9$ \cite{burland:4548,wright:934} while in EPA glass at 77K $k^{(H)}/k^{(D)} = 1.7$ \cite{li:605}.
It should be pointed out that theoretical studies of the isotope effect in the non-radiative decay of benzene have 
shown little agreement with experimental results.  For example,  in Ref. ~\cite{burland:4548}, Burland and Robinson use a 
level counting approach and obtain an estimate of  $k^{(H)}/k^{(D)}  \approx 3.7 \times 10^{5}$.  
Other theoretical work using saddle point methods to numerically evaluate the rate constant in the statistical limit
give ratios of $k^{(H)}/k^{(D)}  = 24$ \cite{Fischer1971392} to $k^{(H)}/k^{(D)} = 1.1 \times 10^{4}$ and
 $ 2.2 \times 10^{2}$ depending upon the parameters for the $e_{2g}$ mode\cite{Nitzan:1973fk}.
Finally, recent work by Zamstein using an {\em ab initio} parameterization and 
a phase-space approach for the non-radiative $^{3}B_{1u}\to\ {^{1}A}_{1g}$ triplet to singlet conversion in benzene and benzene-$d_{6}$ gives a  $k^{(H)}/k^{(D)}  = 3.7 - 7.8$ \cite{zamstein:074304}.

\section{Acknowledgments}
This work was supported in part by the National Science Foundation 
(CHE-0712981) and the Robert A. Welch foundation (E-1337).  The authors wish to acknowledge 
Dr. David Beljonne for performing the AM1/SCI calculations used for the parameterization of the model and 
Prof. Robert Silbey for discussions regarding the non-adiabatic couplings.




\balance


\footnotesize{

\providecommand*{\mcitethebibliography}{\thebibliography}
\csname @ifundefined\endcsname{endmcitethebibliography}
{\let\endmcitethebibliography\endthebibliography}{}

}

\end{document}